\documentclass[prd,twocolumn,showpacs,eqsecnum,amsfonts,amssymb,
amsmath,a4paper,english,superscriptaddress, nofootinbib
]{revtex4}
\usepackage{graphicx,multirow}
\newlength{\inda}
\newlength{\indb}

\def\g{\text{\sl g}}

\def\x{{\mathbf x}}

\def\im{{\rm i}}

\def\det{{\mathrm{det}}}

\def\Box{\kern0.5pt{\lower0.1pt\vbox{\hrule height.5pt width 6.8pt
  \hbox{\vrule width.5pt height6pt \kern6pt \vrule width.3pt}
  \hrule height.3pt width 6.8pt} }\kern1.5pt}

\begin{document}
\setcounter{topnumber}{1}

\title{Quasi-normal mode analysis in BEC acoustic black holes}
\author{C. Barcel\'{o}}
\affiliation{Instituto de Astrof\'{i}sica de Andaluc\'{i}a, CSIC, Camino Bajo de Hu\'{e}tor 50, 18008 Granada, Spain}
\author{A. Cano}
\affiliation{Instituto de Astrof\'{i}sica de Andaluc\'{i}a, CSIC, Camino Bajo de Hu\'{e}tor 50, 18008 Granada, Spain}
\author{L. J. Garay}
\affiliation{Departamento de F\'{i}sica Te\'{o}rica II, Universidad Complutense de Madrid, 28040 Madrid, Spain}
\affiliation{Instituto de Estructura de la Materia, CSIC, Serrano 121, 28006 Madrid, Spain}
\author{G. Jannes}
\affiliation{Instituto de Astrof\'{i}sica de Andaluc\'{i}a, CSIC, Camino Bajo de Hu\'{e}tor 50, 18008 Granada, Spain}
\affiliation{Instituto de Estructura de la Materia, CSIC, Serrano 121, 28006 Madrid, Spain}

\date{\today}

\begin{abstract}

We perform a quasi-normal mode analysis of black hole configurations
in Bose-Einstein condensates (BEC). In this analysis we use the full
Bogoliubov dispersion relation, not just the hydrodynamic or
geometric approximation. We restrict our attention to one-dimensional
flows in BEC with step-like discontinuities. For this case we show
that in the hydrodynamic approximation quasi-normal
modes do not exist. The full dispersion relation, however, allows the existence of quasi-normal modes. Remarkably, the spectrum of these modes is not discrete but continuous.

\end{abstract}
 \maketitle

\section{Introduction}
\label{S:introduction}
Analogue models of gravity \cite{review,analogue-book} allow the
study of certain geometrical aspects of linearized general relativity
(GR). When the analogue models are based on an intrinsically quantum
system, such as Bose-Einstein condensates (BECs) \cite{sonicBH1,
sonicBH2, barcelo-bec}, this opens up the possibility of studying some
corrections to this geometric description. These corrections ---
encoded in the form of non-relativistic or modified dispersion relations, similar to what
is expected in a wide range of scenarios for gravity at Planck-scale
energies, leading to possible violations of Lorentz invariance \cite{lorentz_violations} --- are then of a quantum origin, but
they express themselves in a classical manner. In general, these
corrections have to be taken into account in any treatment involving high energies (high
frequencies). In BECs in particular, in the presence of horizons, the deviations from the geometric or hydrodynamic regime are important
even at low energies \cite{instabilities}.

In a previous article \cite{instabilities}, the stability of several
one-dimensional configurations in BECs was studied. These
configurations were chosen in analogy with the geometries
associated with gravitational black holes.
Here we will elaborate further on black hole
configurations, which were shown to be devoid of instabilities,
and examine their quasi-normal or relaxation modes. The analysis we
carry out is analogous to the standard quasi-normal mode (QNM)
analysis in gravity
\cite{nollert, kokkotas}. According to GR, when black holes are
perturbed, they emit gravitational waves in the following way. The
initial regime depends mainly on the concrete form of the
perturbation. Then, an oscillatory phase is attained in which the
precise form of the emitted wave depends only on the properties of the
black hole itself. Finally, a polynomial tail characterizes the return
to equilibrium. In the intermediate phase, a discrete set of complex
frequencies are excited: the quasi-normal modes of our concern. Since
quasi-normal modes are modes of decay, i.e. of energy dissipation,
they are outgoing. In GR this means that their group velocities are
directed outwards both in the asymptotic region and at the
horizon. Moreover, as we will illustrate, in GR this also
automatically implies that they are non-normalizable or divergent.

A general investigation of quasi-normal modes in analogue black holes
was carried out in \cite{berti}. As we will briefly point out, in standard GR in $(1+1)$ dimensions, quasinormal modes do not exist. In higher dimensions, a discrete spectrum appears \cite{nollert, kokkotas}. Similar results were found in \cite{berti} for analogue black holes. Here the essential element that we add is the influence of a
non-relativistic contribution to the dispersion relation on these
QNMs\footnote{Modifications of the quasi-normal mode
spectrum of black holes due to Lorent violations have also
been investigated in \cite{chen} but from a different perspective. They
introduce spontaneous Lorentz-symmetry breaking
terms in an effective Lagrangian via non-zero expectation
values of different tensor fields in the vacuum, thereby restricting
attention to a modified Dirac equation in a Schwarzschild background.
However, these terms do not lead to modified dispersion relations
of the type analyzed here.}. With the full Bogoliubov dispersion relation in BEC, not only does a
QNM-spectrum appear even in the one-dimensional case studied here, but
this spectrum turns out to be continuous. This result suggests the straightforward speculation that in $(3+1)$ dimensions, for both analogue and GR black holes, there will also be continuous regions of QNMs when modifications of the dispersion relation with respect to the hydrodynamic or relativistic case are taken into account.


The structure of this paper is as follows. In the next two sections we
will review the steps that lead from the general description of a BEC
to the simplified formalism for one-dimensional configurations with
abrupt discontinuities, and highlight the modified dispersion
relation. In Sec.~\ref{S:boundary_conditions} we will discuss the
boundary conditions for our problem, and compare with the standard
quasi-normal mode analysis of gravitational black holes. Then, in
Sec.~\ref{S:discussion} we present and discuss our main results.
%
\section{Preliminaries}
\label{S:preliminaries}
In this section we will briefly review the description of background
BEC configurations and their perturbations, with particular emphasis
on the modified dispersion relation. The main aim here is to fix
notation. General reviews of BECs can be found e.g. in
\cite{dalfovo,castin} while the type of configurations discussed here
were amply described in \cite{instabilities}.

A dilute gas of interacting bosons can be described in terms of quantum field operators $\widehat \Psi({\mathbf x})$ [and $\widehat \Psi^\dag({\mathbf x})$] that annihilate [and create] particles. 
When this gas condensates, $\widehat \Psi$ can be separated into a
macroscopic wave function $\psi$ and a field operator $\widehat
\varphi$ describing quantum fluctuations:
\mbox{${\widehat \Psi}=\psi+{\widehat \varphi}$}.
The macroscopic wave
function $\psi$ satisfies the Gross-Pitaevskii (GP) equation
\begin{align}
 \im \hbar \; \frac{\partial }{\partial t} \psi(t,\x)= \left(
 - \frac{\hbar^2}{2m} \nabla^2
 + V_{\rm ext}(\x)
 + \g \; |\psi|^2 \right) \psi(t,\x),
\label{GP}
\end{align}
where $m$ is the boson mass, $V_\text{ext}$ the external potential
and $\text{\sl g}$ a coupling constant which is related to the
corresponding scattering length $a$ through $\g ={4\pi \hbar^2 a
/m}$.

This equation can be expressed in terms of hydrodynamic quantities
such as the local speed of sound $c$ and the velocity of the fluid
flow $\mathbf v$.  This proceeds by first introducing the Madelung
representation for the order parameter in the GP equation:
\begin{eqnarray}
\psi = \sqrt{n}e^{\im \theta/\hbar} e^{- \im \mu t/\hbar }.
\label{madelung}
\end{eqnarray}
Here $n$ is the condensate density, $\mu$ the chemical
potential and $\theta$ a phase factor.
Next, as we are interested in the acoustic perturbations of the
BEC, we linearize in $n$ and $\theta$:
\begin{subequations}
\begin{align}
n (\x, t) &= n_0(\x) + \g^{-1} \widetilde n_1(\x, t),
\\
\theta(\x, t) &= \theta_0(\x) + \theta_1(\x, t).
\end{align}
\end{subequations}
Then, the following set of equations for the linear
perturbations is obtained from the complex GP equation
\begin{subequations}
\label{GP_lin}
\begin{align}
\partial _t \widetilde n_1 &= - \nabla \cdot
\left( \widetilde n_1 \mathbf v + c^2 \nabla \theta_1 \right),
\label{GP1_lin}
\\
\partial _t \theta_1
&= -\mathbf v \cdot \nabla \theta_1 - \widetilde n_1 +\frac{1}{4} \xi^2
\nabla \cdot \left[c^2 \nabla \left( \frac{\widetilde n_1}{c^2}\right)\right],
\label{GP2_lin}
\end{align}
\end{subequations}
which are equivalent to the Bogoliubov equations, and where $\xi \equiv \frac{\hbar}{mc}$ is the healing length, and $c$ and $\mathbf v$ are
\begin{equation}
c^2 \equiv \g n_0/m, \qquad \mathbf v \equiv \nabla \theta_0 /m.
\end{equation}
The last term on the right-hand side in \eqref{GP2_lin} is the
so-called ``quantum potential'' term.

At this point, let us briefly recall that in the hydrodynamic limit, which is obtained by neglecting the quantum potential term just mentioned, an effective Lorentzian metric can be constructed of
the type:
\begin{align}
(g_{\mu \nu}) \propto
\begin{pmatrix}
v^2-c^2 && -\mathbf v^\text{T}\\
-\mathbf v && \mathbf \openone
\end{pmatrix},
\label{metric}
\end{align}
which historically formed the onset of the interest in hydrodynamic analogue models for gravity (for an overview, see \cite{review} and references therein).

From now on, consider one-dimensional profiles. The equations \eqref{GP_lin} can be solved in a homogeneous region ($c$ and $v$ constant) by writing the perturbations in plane-wave form
\begin{subequations}\label{plane-waves}\begin{align}
\widetilde n_1(x,t)&=
A e^{i(k x - \omega t)}, \\
\theta_1(x,t)&= B e^{i(k x - \omega t)}.
\end{align}\end{subequations}
This immediately leads to the local dispersion relation\footnote{The term `dispersion relation' is sometimes reserved strictly for an equation of the form $\omega=\omega(k)$, while an equation such as \eqref{dispersion_xi} would then have to be called e.g. the `characteristic equation'. For the sake of clarity, we will always use the term `dispersion relation' to refer to an equation of the form \eqref{dispersion_xi} or \eqref{quadr_dispersion}.}
\begin{align}\label{dispersion_xi}
(\omega - vk)^2= c^2k^2 + \frac{1}{4} c^2 \xi^2 k^4.
\end{align}
This dispersion relation is \textit{quartic} in $k$. It should be compared to the hydrodynamic dispersion relation that is obtained by neglecting the quantum potential:
\begin{align}\label{quadr_dispersion}
(\omega - vk)^2= c^2k^2.
\end{align}
This hydrodynamic dispersion relation is of the usual \textit{quadratic}
form.

However, here we are not so much interested in this hydrodynamic limit, but precisely in deviations from it. Therefore we will always work with the full dispersion relation, and only use the hydrodynamic limit as a means of comparison. The key point is that, because the full dispersion relation is quartic, when writing the mode corresponding to a certain frequency $\omega$:
\begin{equation}\label{bogoliubov_modes}
u_\omega=e^{-i\omega t}\sum_j A_j e^{ik_jx},
\end{equation}
there are now four contributions to this $u_\omega$-mode, stemming from the four values of $k$ associated to each value of $\omega$. In the hydrodynamic limit there are obviously only two contributions to each $u_\omega$-mode:
\begin{equation}
k_{1,2} = \omega/(v \pm c).
\end{equation}
In the configurations analyzed in this paper, the additional,
``non-hydrodynamic'' modes will turn out to play a crucial role even
at low frequencies.
%
\section{Background configurations and numerical method}
\label{S:numerical}
\begin{figure}[tbp]
\includegraphics[width=.4\textwidth,clip]{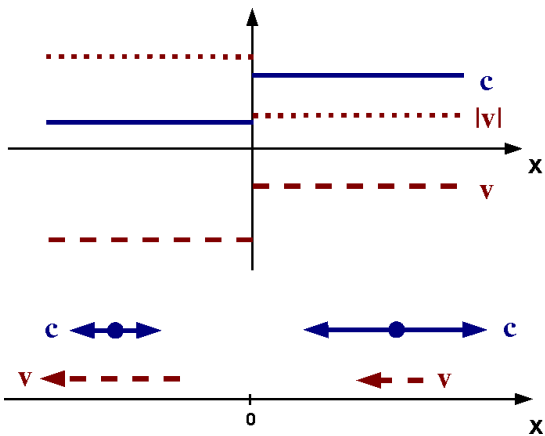}
\bigskip
\caption{(Color online) Flow and sound velocity profile consisting of two homogeneous regions with a step-like discontinuity at $x=0$ simulating a black hole-like configuration in a BEC. The solid blue line represents the speed of sound $c$, the dashed red line the fluid velocity $v$. In the upper part of the picture, the negative value of $v$ indicates that the fluid is left-moving. For $x>0$, the fluid is subsonic since $c>\lvert v \rvert$. At $x<0$ it has become supersonic. At $x=0$, there is a sonic horizon.}
\label{F:profiles}
\end{figure}

We are looking for quasi-normal modes of black hole configurations in
one-dimensional BECs of the type represented in fig.~\ref{F:profiles}. These profiles are piecewise homogeneous with a single discontinuity. In order to find these relaxation modes, first, we will need a way to describe the mode solutions that
characterize the profiles under study. The idealized form of these profiles
allows a simplified description in each homogeneous region. Then, the
various sections have to be linked through matching conditions at the
discontinuity. After finding all the mode solutions, the quasi-normal modes will be those that moreover satisfy the adequate boundary conditions. We will describe the matching conditions in the current section, together with a straightforward algorithm for
numerical implementation. The boundary conditions will be the topic of
the next section.
%

Integrating \eqref{GP_lin} in an infinitesimal neighbourhood of the discontinuity (say $x=0$) and simplifying with the use of the continuity equation ($vc^2 = \text{constant}$) leads, for our choice of a piecewise homogeneous background, to the following set of matching conditions \cite{instabilities}:
\begin{subequations}\label{matching}\begin{align}
[\theta_1]=0,\qquad [c^2 \partial_x \theta_1]=0,\label{matching_a}\\
\left[ \frac{\widetilde n_1}{c^2}\right]=0,
\qquad \left[\partial_x
\widetilde n_1\right]=0,
\end{align}\end{subequations}
where the brackets indicate, for example, \mbox{$[\theta_1]=\left.\theta_1\right|_{x=0^+} -\left.\theta_1\right|_{x=0^-}$}.

In each homogeneous section, we look for a solution for a particular frequency $\omega$ as a sum of $k$-modes of the form
\begin{align}\label{general_solution}
\widetilde n_1 =&
\displaystyle\sum _j A_j e^{i(k_jx - \omega t)},\\
\theta_1 =&
\displaystyle\sum _j A_j \frac{\omega - vk_j}{ic^2k_j^2} e^{i(k_jx- \omega t)}.
\end{align}
where the sum is over the four $k_j$-values associated to $\omega$ in that homogeneous region.

The conditions \eqref{matching} can be written in matrix form
\begin{equation}
\Lambda_{ij}A_j=0.
\end{equation}
Since there are eight free parameters $A_j$ for each discontinuity (four in each region),
and only four conditions, matching is in principle always
possible. However, there are additional conditions: the boundary
conditions which we will discuss in the next section. These boundary
conditions can be added to the matrix $\Lambda_{ij}$, to obtain a
$(4+N)\times 8$ matrix $\widetilde\Lambda_{ij}$, with $N$ the number
of constraints on the coefficients $A_j$ that follow from the boundary
conditions. Depending on the value of $N$, we can have the following
cases.
\begin{itemize}
\item If $N<4$, then there will always exist a non-trivial solution ${A_j}$, since there are more degrees of freedom than constraints. Therefore, a region of the complex $\omega$-plane where $N<4$ will represent a continuous region of eigenfrequencies.
\item If $N=4$, then we have a $8 \times 8$ system of equations and there is a non-trivial solution ${A_j}$ only if $\det(\widetilde\Lambda)=0$. Generally speaking, this means that in a region where $N=4$, we might at the very most expect isolated discrete eigenfrequencies, but no continuous zones of solutions.
\item If $N>4$ then the system can be split into two (or more) $8\times 8$ subsystems, each with a subdeterminant $\lambda_i$. For a non-trivial solution to exist, each of these subsystems should fulfill the previous condition $\det(\lambda_i)=0$.
\end{itemize}
To summarize, we can define a non-negative function $F(\omega)$ such that
\begin{itemize}
\item $F(\omega)=0$ if $N<4$,
\item $F(\omega)=|\det(\widetilde\Lambda)|$ if $N=4$,
\item $F(\omega)=\sum|\det(\lambda_i)|$ if $N>4$.
\end{itemize}
The eigenfrequencies of the system are those values of $\omega$ for which $F(\omega)=0$, and since the quasi-normal modes are relaxation modes they will be located in the lower half complex $\omega$ plane.
%
\section{Quasi-normal mode analysis and boundary conditions}
\label{S:boundary_conditions}
The essential ingredient that is still missing to calculate the quasi-normal modes in these BEC black hole configurations are the boundary conditions. We will first take a look at the boundary conditions that are used to calculate QNMs in GR. This can be immediately extrapolated to BECs in the hydrodynamic limit. However, as we will show, in this hydrodynamic limit in $(1+1)$ dimensions, there exist no QNMs. Finally, we will come to the boundary conditions with the full dispersion relation.
%
\subsection{Quasinormal modes of a gravitational black hole}
\label{SS:QNM_gravity}
Quasi-normal modes are simply the modes of energy dissipation of a perturbed object.
Gravitational black hole QNMs in particular are the relaxation modes that characterize the pulsations of the black hole after perturbations initiated internally in the spacetime surrounding the black hole.

The fact that these modes decay in time means that they can be represented by complex frequencies $\omega$ with Im$(\omega) < 0$. The fact that they are relaxation modes means that they will have to be outgoing in the asymptotic region. At the horizon, QNMs are also required to be outgoing because in GR nothing can escape through the horizon (note that in QNM terminology ``outgoing'' means directed towards the exterior of the spacetime region connected to an asymptotic observer, in other words at the horizon this means directed towards the singularity). So the boundary conditions that are imposed in GR to find the QNMs of a black hole are simply that these modes should be outgoing both in the asymptotic region and at the horizon.

Since these QNMs are outgoing, they are also divergent. Indeed, because of the quadratic dispersion relation, say $\omega^2=c^2k^2$, we have for the group velocity $v_g$ and for the imaginary part of $k$:
\[
v_g=\pm c, \qquad \text{Im}(k)=\text{Im}(\omega) /v_g
\]
and therefore, since QNMs have Im$(\omega)<0$:
\[
\text{sign}[\text{Im}(k)]=-\text{sign}[v_g].
\]
To sum up, QNMs in a gravitational black hole are found by imposing the boundary condition that they should be outgoing, and because of the quadratic dispersion relation they are also automatically divergent.

\subsection{BEC hydrodynamic limit: absence of quasi-normal modes in one-dimensional flows}
\label{SS:hydrodynamic_QNM}

In BECs in the hydrodynamic limit, decaying outgoing modes are also automatically divergent, again because of the quadratic dispersion relation. Indeed,
\[
v_g=v \pm c,\qquad
\text{Im}(k)=\text{Im}(\omega)/v_g
\]
and therefore $\text{sign}[\text{Im}(k)]=-\text{sign}[v_g]$ as before. Let us see whether QNMs can exist in this case.

In the hydrodynamic limit, the quantum potential term in \eqref{GP_lin} is neglected, and so upon integration of the remaining terms in these equations, only two matching conditions remain:
\begin{equation}
[\theta_1]=0, \qquad [v \widetilde n_1 + c^2\partial_x \theta_1]=0.
\end{equation}
Note that this second equation cannot be simplified as in \eqref{matching}, because this requires the use of a third matching condition which came precisely from integration of the quantum potential.

The hydrodynamic solution for a particular \mbox{frequency $\omega$},
\begin{align}
\widetilde n_1 &= A_1 e^{i(k_1x - \omega t)} + A_2 e^{i(k_2x - \omega t)},\\
\theta_1 &= B_1 e^{i(k_1x- \omega t)} + B_2 e^{i(k_2x- \omega t)},
\end{align}
has to satisfy $A_j = i(\omega-vk_j)B_j$.  Assume that the configuration is completely subsonic. Then, for $x<0$, \mbox{$k_1=\omega/(v+c)\equiv k_\text{in}^-$} is \textit{in}going, whereas \mbox{$k_2=\omega/(v-c)\equiv k_\text{out}^-$} is \textit{out}going, and vice versa for $x>0$, where we write $k_\text{in}^+$ and $k_\text{out}^+$ respectively. In terms of the associated
coefficients $B_j$, the matching conditions then become
\begin{align}
B_\text{out}^+ + B_\text{in}^+ &= B_\text{in}^- + B_\text{out}^-,\\
c^+(B_\text{out}^+ - B_\text{in}^+) &= c^-(B_\text{in}^- - B_\text{out}^-),
\end{align}
or in matrix form
\begin{align}
\begin{pmatrix}
B_\text{in}^+\\B_\text{out}^+
\end{pmatrix}
=
\frac{1}{2c^+}
\begin{pmatrix}
c^+ - c^- && c^+ + c^-\\
c^+ + c^- && c^+ - c^-
\end{pmatrix}
\begin{pmatrix}
B_\text{in}^-\\B_\text{out}^-
\end{pmatrix}.
\label{matrix}
\end{align}
For a solution to be outgoing in both regions, 
both $B_\text{in}^+$ and $B_\text{in}^-$ should vanish. This means that the coefficient $c^+ + c^-$ connecting $B_\text{in}^+$ with $B_\text{out}^-$ must be zero. But since $c$ is always a positive integer, this is impossible. Therefore, there are no quasi-normal solutions in the hydrodynamic limit.


The above calculation is strictly valid only for completely
subsonic profiles. When the left-hand side is supersonic, the
hydrodynamic approximation breaks down in the passage through the
sonic horizon \cite{instabilities}. Both modes in the supersonic region become outgoing at $x\rightarrow -\infty$, while there is still one outgoing mode in the right asymptotic (subsonic) region. Since the matrix in \eqref{matrix} is well defined, it seems then that some QNMs exist for these configurations. However, this is a spurious effect caused by the subsonic to supersonic step-like discontinuity in combination with the hydrodynamic regime. Precisely because of the presence of the horizon, we know that none of the modes at the left-hand side can be connected to the outgoing mode at the right-hand side. So QNMs cannot exist. From a point of view more similar to GR, when the $(x<0)$ region is supersonic, it becomes disconnected from the subsonic region $(x>0)$. But then, just like in GR, the requirement that QNMs must be outgoing in the left asymptotic region should be replaced by the requirement of being outgoing at the horizon. Because of the homogeneity of the profile, none of the modes can be outgoing both at the right asymptotic region and at the horizon. So in any case, QNMs cannot exist in the hydrodynamic limit in $(1+1)$ dimensions, regardless of the sub- or supersonic character of the profile.

Note that in GR in $(1+1)$ dimensions, a more general argument prohibits
the existence of QNMs. Indeed, all two-dimensional metrics are
conformally flat. In addition, the d'Alembertian equation in two
dimensions is conformally invariant \cite{birrell-davies}. Therefore,
all the solutions of the system are conformally equivalent to plane waves, and so they don't satisfy the requirement of being outgoing both in the asymptotic region and at the horizon. Hence QNMs cannot exist in $(1+1)$ GR. The situation is completely different in $(3+1)$ dimensions, in which GR black holes exhibit a discrete QNM spectrum.

%
\subsection{Boundary conditions with the full dispersion relation}
\label{SS:boundary_conditions}
As we just discussed, in BECs in the hydrodynamic limit, decaying
outgoing modes are automatically divergent. But when looking at the
full dispersion law, this relation is no longer valid. Note that for
our choice of profiles, we have two asymptotic regions, and moreover
no prohibition of crossing the horizon in any direction, since
the permeability of the horizon is an essential
feature of modified dispersion relations \cite{instabilities}. An
outgoing mode is then one that has a group velocity $v_g<0$ for $x<0$ and $v_g>0$ for $x>0$, where $v_g$ is obtained from the full dispersion
relation:
\begin{equation}\label{group-velocity}
v_g(\omega, k) \equiv \text{Re} \left( \frac{d \omega}{dk}\right) =
\text{Re}\left(
\frac{c^2 k + {\frac{1}{2}} \xi^2 c^2 k^3}{\omega - vk} + v
\right).
\end{equation}
Clearly, there is no immediate connection with the sign of Im$(k)$ anymore, so the automatic divergent character of outgoing modes is lost.

In any case, quasi-normal modes should be outgoing, and this remains the essential boundary condition. Additionally, one might wonder whether the QNMs will be convergent, or divergent after all (in spite of the modification of the dispersion relation). An easy way to check this is by applying convergence as an additional boundary condition and comparing the resulting spectrum with the one obtained without this additional condition.

To sum up, quasi-normal modes should be outgoing. In GR and in the
hydrodynamic approximation, this automatically implies a
divergent behaviour in the asymptotic regions.
However, in both these cases, there are no QNMs in $(1+1)$ dimensions
(as opposed to $(3+1)$ GR black holes with their discrete QNM
spectrum). In BECs with the full modified dispersion relation,
outgoing modes are no longer automatically divergent. The boundary
condition to determine QNMs in the $(1+1)$ BEC black hole
configurations under study is therefore that they should be
outgoing. Nonetheless, we will check whether they are convergent or
divergent.
%
\section{Results and discussion}
\label{S:discussion}
In this section we present and discuss results for the quasi-normal mode spectrum of black holes in one-dimensional BECs obtained with the full dispersion relations.

\begin{figure}[tbp]
\includegraphics[width=0.4\textwidth,clip]{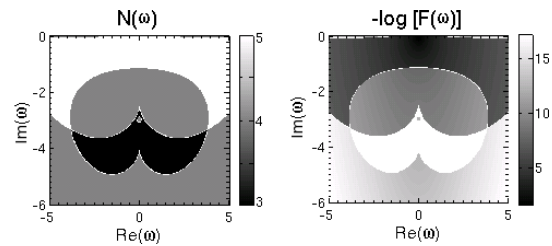}
\bigskip
\caption{Plots illustrating the quasi-normal mode spectrum for a one-dimensional black hole configuration in BEC. The left-hand part of the picture represents the number $N$ of constraints following from the boundary condition that QNMs should be outgoing. The right-hand part shows the function log$[F(\omega)]$ defined in Sec.~\ref{S:numerical}, where $F(\omega)=0$ corresponds to quasi-normal modes. The QNM spectrum consists of the continuous region where $N<4$. [The numerical values used for these plots, in units such that the healing length $\xi=1$, are $c=1; v=0.7$ in the subsonic region and $v=1.8$ in the supersonic region.]}
\label{F:QNM_outgoing}
\end{figure}

Fig.\ref{F:QNM_outgoing} illustrates the QNM spectrum for an acoustic black hole containing a single discontinuity, subsonic at the right-hand side and supersonic at the left. Remember from \cite{instabilities} that such a black hole did not possess unstable eigenfrequencies, so it indeed makes sense to look for stable modes.

As discussed in the previous section, the boundary condition that determines the existence of QNMs is that they should be outgoing in both asymptotic regions. Surprisingly, as can be seen from Fig.\ref{F:QNM_outgoing}, not only does such a QNM spectrum indeed show up, but this spectrum is actually continuous.
%

\begin{figure}[tbp]
\includegraphics[width=0.4\textwidth,clip]{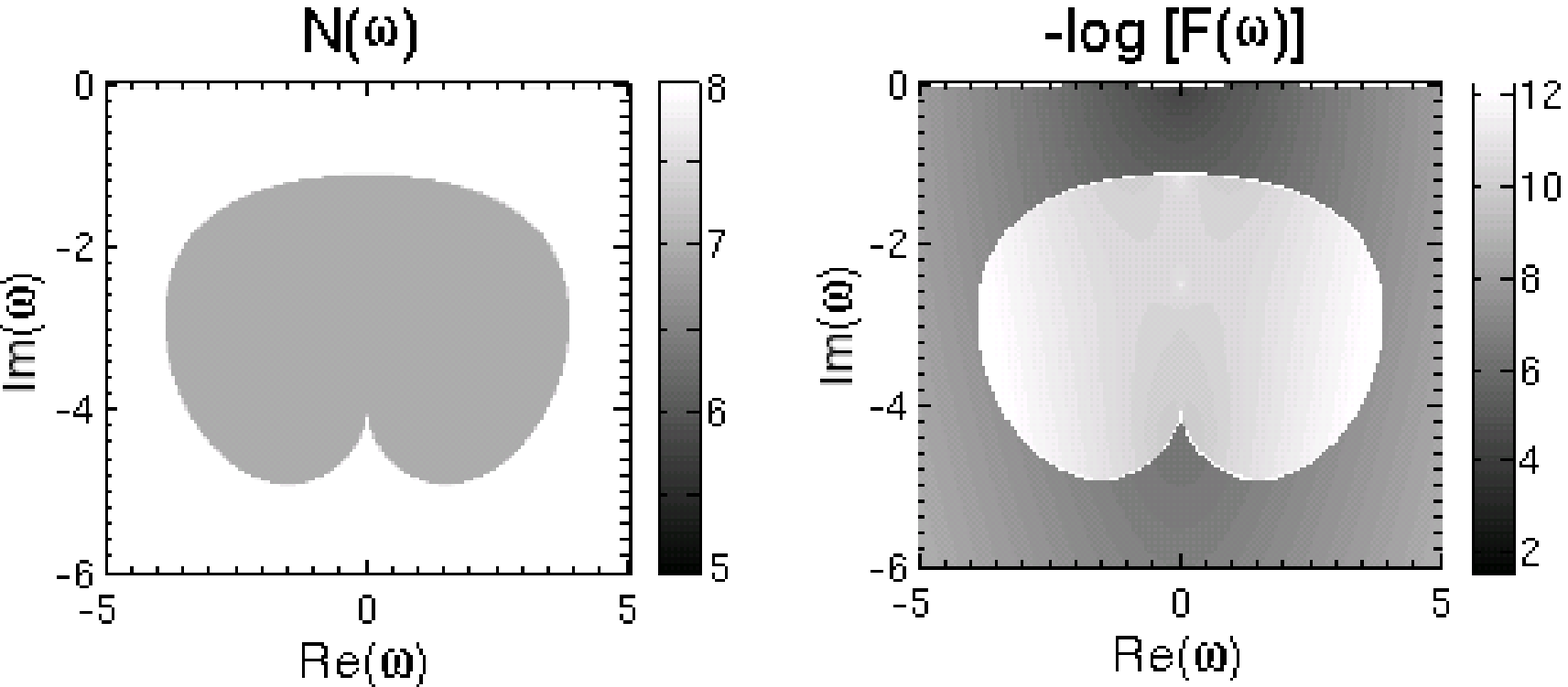}
\includegraphics[width=0.4\textwidth,clip]{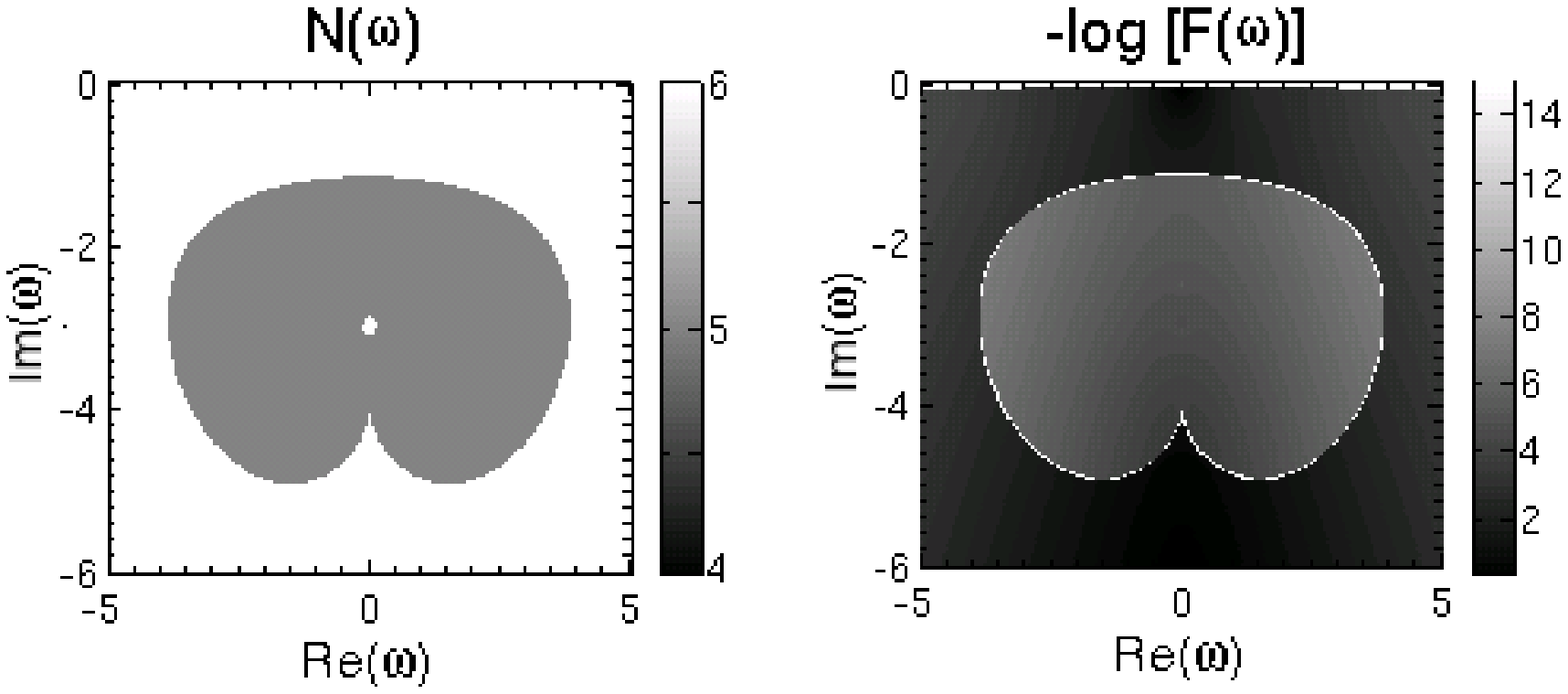}
\includegraphics[width=0.4\textwidth,clip]{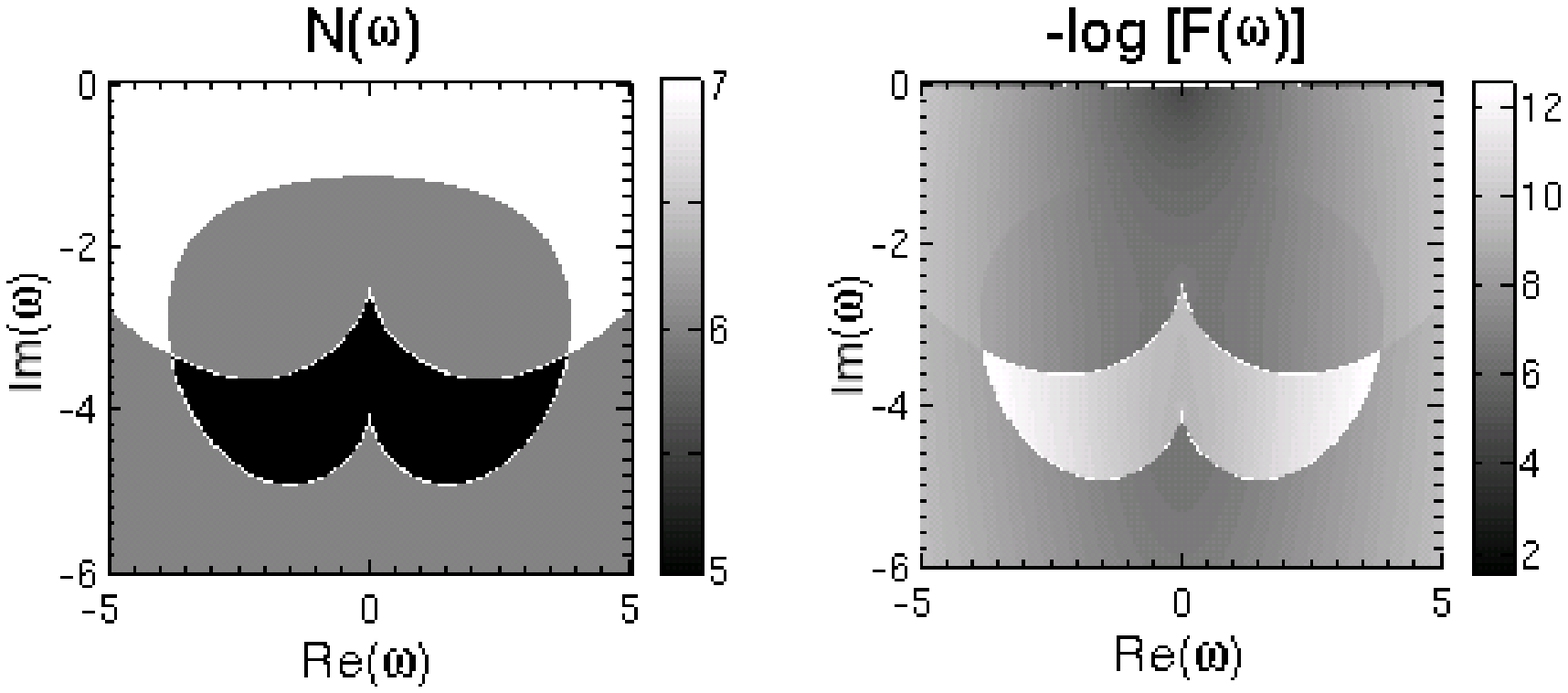}
\bigskip
\caption{Alteration of QNM spectrum resulting from imposing the additional boundary conditions of (from top to bottom) convergence in both asymptotic regions, in the right asymptotic region only, and in the left asymptotic region only. Since $F(\omega)$ never reaches zero, it is as if the QNM spectrum of Fig.~\ref{F:QNM_outgoing} had disappeared. In other words, the QNMs are divergent in both asymptotic regions. [Same numerical values for $c$ and $v$ as in Fig.~\ref{F:QNM_outgoing}.]}
\label{F:QNM_convergence}
\end{figure}

%
%
The next question is whether these modes are convergent or not. In Fig.\ref{F:QNM_convergence}, we have imposed convergence, respectively at both sides, in the right asymptotic region only, and in the left asymptotic region (the singularity) only.
%
%
The QNM spectrum disappears as soon as the convergence condition is imposed at either side. In other words, the \textit{quasi-normal modes are divergent}. Remember from the previous section that for outgoing hydrodynamic modes, this divergent character is automatic. However, we have also seen that when only retaining the hydrodynamic modes, there are no QNMs. So although the QNMs that we have just found are divergent, the inclusion of the non-hydrodynamic modes is essential for their existence, and so their divergent character is not ``automatic'', i.e. it is not due to a strict relation between the signs of $v_g$ and Im$(k)$. As a matter of fact, we have checked that the QNMs disappear when the convergent $k$-contributions are excluded.

We have also calculated the QNM spectrum for other black-hole like configurations that were shown to be devoid of instabilities in \cite{instabilities}, and we have checked that a similar continuous spectrum of QNMs appeared in all of them. This indicates that the continuous character of the QNM spectrum is an essential consequence of the modified dispersion relation, and not just of the particular characteristics of a precise configuration.

To sum up, there is a continuous region of \mbox{frequencies $\omega$} such that they have associated modes
\[
u_\omega=e^{-i\omega t}\sum_{j=1}^4 A_j e^{ik_jx},
\]
which are outgoing at both ends. These modes are divergent, just like in the hydrodynamic case, although the fact that here non-hydrodynamic modes come into play means that this relation cannot be derived directly from the dispersion relation.
%
\section{Conclusions}
We have examined the quasi-normal modes of one-dimensional black hole configurations in BECs, and laid particular emphasis on the importance of the deviation of the full dispersion relation with respect to a relativistic or acoustic one. The full dispersion relation is quartic in BECs, as compared to the usual second order of its hydrodynamic and general relativistic equivalent. With such quadratic dispersion relations, outgoing relaxation modes are automatically divergent. In particular, this argument is valid for QNMs in GR black holes. But the modification of the dispersion relation means that this relation is no longer automatic. Nevertheless, the quasi-normal modes in the systems discussed here are also divergent.

More importantly, the QNM spectrum that was obtained in these systems with a modified dispersion relation consists of a continuous region in the complex frequency plane. The importance of this result is easy to see when taking into account that both in GR and in BECs in the hydrodynamic limit, QNMs simply do not exist in $(1+1)$ dimensions, while in higher dimensions the quasi-normal modes of a GR black hole form a discrete spectrum. So due to the modification of the dispersion relation, the QNM spectrum in one-dimensional BEC black holes changes from non-existent to a continuous region of frequencies.

We would like to end this paper with the straightforward
speculation that the discrete spectrum which is obtained in
the usual quasi-normal mode analysis in standard $(3+1)$ GR, will
also develop continuous regions when taking the modification of the
dispersion relation at high energies into account. The existence of these
continuous bands could become a signal of trans-Planckian physics encoded in the
emitted spectrum of gravitational waves.
\acknowledgements
C.B. has been funded by the Spanish MEC under project
FIS2005-05736-C03-01 with a partial FEDER contribution. G.J. was
supported by CSIC grant I3P-BPD2005 of the I3P
programme, cofinanced by the European Social Fund,
and by the Spanish MEC under project FIS2005-05736-C03-01.
L.G. was supported by the Spanish MEC under project FIS2005-05736-C03-02.
The authors also acknowledge support from the Spanish MEC project FIS2006-26387-E.



\end{document}